\documentstyle[multicol,aps]{revtex}
\begin{document}
\draft
\title{Baryon Isocurvature Perturbation in the Affleck-Dine
  Baryogenesis mechanism}
\author{Kazuya Koyama $^1$  and Jiro Soda $^2$}
\address{$^1$ Graduate School of Human and Environmental Studies,Kyoto 
University,Kyoto 606-8501, Japan \\
$^2$ Department of Fundamental Physics, FIHS Kyoto University, Kyoto 606-8501,
 Japan}
\date{February 11.1998}
\maketitle

\begin{abstract}
 We propose a primeval baryon isocurvature model in the context of the 
Affleck-Dine (AD) baryogenesis mechanism. 
Quantum fluctuations of the AD field during 
inflation produce fluctuations in the baryon number. We obtain a model which 
gives both the appropriate baryon to entropy ratio of the order $10^{-11}$ 
and the gaussian baryon isocurvature perturbation of the order $10^{-3}$ at 
the break of the spectrum $1 \mbox{Mpc}$ with the steep spectrum $n=-1.5$ on 
large scales $>1 \mbox{Mpc}$.
\end{abstract}
\pacs{PACS numbers: 98.80.Cq, 11.30.Fs, 95.35.+d}

\begin{multicols}{2}
The primeval baryon isocurvature model (PBI) which attempts to account for 
the large scale structure of the universe only by the baryon and the 
radiation has several appealing properties which are lacking in standard CDM 
model \cite{Peebles1}. Under the assumption that there exist primeval 
fluctuations in the baryon to entropy ratio with a steep spectrum on large 
scales, PBI fits two sets of observations \cite{Peebles2,Cen}. First, large 
power on small scales in the fluctuations spectrum produces galaxy at a 
relatively early stage, which is supported by recent galaxy surveys. Second, 
the large scale correlations in the distributions of galaxies would be 
produced at matter-radiation Jeans scale $\sim 100$Mpc. 

So far, most of studies about PBI are started by assuming the existence of 
ad hoc primordial fluctuations. To realize PBI in the physics of early 
universe, one needs a model of baryogenesis which can give both the 
appropriate baryon to entropy ratio $n_B/s$ of the order $10^{-11}$ and 
its fluctuations with the steep spectrum. The tilted spectrum is necessary 
to suppress the cosmic microwave background (CMB) anisotropy still keeping 
sufficient amplitude on the galaxy scale \cite{Efstathiou}. A possible model 
is the one based on the grand unified theory (GUT) baryogenesis and inflation.
 In Ref \cite{Yokoyama}, the baryogenesis from the decay of the heavy Majorana 
lepton with both hard and soft CP violation is considered and the non-gaussian
 fluctuations with the steep spectrum are obtained. Generally, however, the
 GUT scale ($M_G$) inflation produces the adiabatic perturbation of the order
 $(M_G/M_p)^2 \sim 10^{-5}$, where $M_p$ is the planck scale, so one needs 
the artificial suppression to ensure it's subdominance to the isocurvature 
perturbation. If one takes the lower energy as the scale of the inflation, the
 adiabatic perturbation is suppressed, but the reheating temperature also 
decreases, so the GUT baryogenesis would not work well. Then, we need the 
model of baryogenesis at lower energy  scales. Unfortunately, the weak scale 
baryogenesis in the standard model (SM) of the particle physics is unable to 
produce the sufficient baryon asymmetry \cite{Riotto}, but a viable 
alternative exists in supersymmetric extension of SM, i.e.,the Affleck-Dine 
(AD) mechanism \cite{Affleck,Dine}. The AD mechanism produces the baryon asymmetry efficiently and it is known that scale invariant fluctuations in the 
baryon number are induced by the quantum fluctuations of the field responsible 
for the baryogenesis (AD field) during inflation \cite{iso}. However, as 
mentioned above, the steep spectrum is crucial to realize PBI. 
In this letter, we will show the tilted spectrum of fluctuations is 
implemented in AD baryogenesis mechanism  
by coupling the AD field to another field \cite{Peebles3}.  
 
In the AD mechanism, the baryogenesis occurs from the coherent 
condensation of the AD field, which carries $B - L$ $U(1)$ charge, where $B$ 
and $L$ are the baryon number and the lepton number. The potential of the AD 
field is given by \cite{Dine}
\begin{eqnarray}
V_{\phi}(\phi)&=&(m_{3/2}^2-cH^2) \vert \phi \vert ^2+\left( \frac{\lambda 
(A m_{3/2}+a H)}{n M^{n-3}}\phi^n+(h.c) \right) \nonumber\\
&& \quad + \vert \lambda \vert ^2 \frac{\vert \phi \vert ^{2n-2}}{M^{2n-6}} ,
\label{eqn:b3}
\end{eqnarray} 
where $m_{3/2}$ is the gravitino mass, $H$ is the Hubble parameter, $M$ is 
some large mass scale, $\lambda$ is the dimensionless coupling constant, $n$ 
is some integer larger than $4$, $a=\bar{a}e^{i \theta_a}$ and 
$A=\bar{A} e^{i \theta_A}$ are some complex numbers, and $c$ is a real 
constant. Here we will consider the case $n=4$ and $\lambda=1$
\cite{Exam}. The important 
point of the model of Ref \cite{Dine} is that the negative effective mass 
term is assumed to exist during inflation. During inflation,
the AD field has large value and subsequent to inflation, it spirals into 
the origin exciting the baryon number and decays to the baryon after reheating.
Following, we will describe how baryon isocurvature fluctuations with
steep spectrum can be obtained in this model of bayogenesis.

(1){\it During inflation} ($H \gg m_{3/2}$), we can ignore the terms 
proportional
 to $m_{3/2}$. Due to the negative mass term proportional to $H$, the 
potential has global minimum far from the origin and undergoes quantum 
fluctuations of the order $H$ about the minimum.\cite{Dine}.
 We consider the scales which are still out of horizon when the 
AD field decays and creates baryons, so the spectrum shape of baryon number
 fluctuations is determined by that of the AD field during inflation. We 
want the tilted spectrum favor to small scales in order to suppress the CMB 
anisotropy \cite{Efstathiou}. Generally, the tilt of the spectrum can be 
obtained if the modes of the fluctuations have the mass proportional to the 
Hubble parameter. This is because amplitudes of modes which are massive 
decrease with time even after they cross the horizon. For example, if the 
mode has the mass $m^2=m_{\ast}^2 H_I^2$ and the Hubble parameter $H_I$ 
during inflation is constant, the mode amplitudes decrease as 
$R^{-3/2+\nu}\:(\nu=(9/4-m_{\ast}^2)^{1/2})$ after the mode crosses the 
horizon. The mode amplitudes are $H_I$ when it crosses the horizon, so at 
the end of the inflation, the mode function becomes 
\begin{equation}
 \delta \bar{\phi_k} \sim \frac{H_I}{\sqrt{2k^3}} \left( \frac{R_k}{R_e} 
\right)^{\frac{3}{2}-\nu} = \frac{H_I}{\sqrt{2k^3}} \left( \frac{k}{R_e H_I} 
\right) ^{\frac{3}{2}-\nu},
\end{equation} 
where $R_k$ and $R_e$ are scale factors when the mode exits the horizon and 
the inflation ends. The tilt from the scale invariant spectrum $n=-3$ is 
given by $\triangle n =3 -2 \nu $, which is larger as the mass becomes large. 
This could produce the isocurvature perturbation with the tilted spectrum 
\cite{Linde}. However, since the comoving scale of galaxies is much larger
 than the horizon scale at the inflationary epoch ($k_{galaxy}/R_e \ll H_I$), 
the fluctuation amplitude on the galaxy scale suffers from severe 
suppression \cite{Yokoyama}. To avoid this problem we must change the break 
of the spectrum from $H_I$ to another scale. This can be done if $\phi$ 
couples to the other field $\psi$ which gives the mass $m^2 \propto \psi^2$ 
to $\phi$. First, we assume the power-low inflation in order 
to obtain the desired behavior of $\psi$ \cite{Peebles3}. The potential of 
the inflaton $\chi$ is
\begin{equation}
U(\chi)=\Lambda^4 \exp \left( - \sqrt{\frac{16 \pi \epsilon}{M_p^2}}
  \chi \right),
\end{equation}
where $\Lambda$ is some mass scale and $\epsilon$, which is less than unity, 
determines the expansion rate during inflation. 
The scale factor and the Hubble parameter are 
$R \propto t^{1/\epsilon},\: H=1/(\epsilon t)$. Second, we couple the AD 
field to the other real field $\psi$ and assume the negative coupling. 
The potential is
\begin{eqnarray}
V(\phi,\psi)&=& V_{\phi}(\phi)+\frac{1}{4}\beta \psi^4 +\frac{1}{2}
\mu^2 \psi^2 -\gamma \psi^2 (\phi^2+\phi^{\ast 2}) ,
\label{1}
\end{eqnarray}    
where $\gamma$ and $\beta$ are some positive coupling constants which we 
assume tiny ($\sim m_{3/2}/M$) and $\beta \psi^2,\mu^2 \gg \gamma
|\phi|^2$ \cite{super}. In this case, we can ignore the effect of 
coupling to $\phi$ in the evolution of $\psi$. Initially, $\psi$ is assumed 
to have the large value then at early times the self-coupling term is 
dominant, and $\psi$ slowly decreases with $t$ ($\propto t^{-1}$) so evolves 
like $H$ ($\psi=(\epsilon/\beta^{1/2})(3 \epsilon^{-1}-2)^{1/2}H$). The 
behavior of $\psi$ changes at $t_1$ when the mass term and the self-coupling 
term become comparable. At that time $\psi=\mu \beta^{-1/2}$, so $t_1$ is 
determined by the mass $\mu$ ($t_1 = \mu^{-1} \epsilon^{-1/2}$). Later, the 
mass term becomes dominant and the amplitude of $\psi$ decays rapidly 
($\psi \propto t^{-3/2 \epsilon}$) \cite{Peebles3}. 
Following, for simplicity, we assume $\bar{a}=0$.  For $t < t_1$, 
due to the coupling with $\psi$, 
the minimum of the AD field $\phi=\bar{\phi}e^{i \theta}$ and the mass 
around the minimum $m^2=m_{\ast}^2 H^2$ is given by
\begin{eqnarray}
\theta_{0 eff} &=& 0,m_{\ast \theta eff}^2 = 4 \epsilon(3-2 \epsilon)
(\gamma/\beta),\nonumber\\
\bar{\phi}_{0 eff} &=& \left( \frac{\sqrt{c_{eff}}H  M}{\sqrt{3}}\right)
^{\frac{1}{2}},
m_{\ast \phi eff}^2 = 4c_{eff},
\label{eqn:b1}
\end{eqnarray} 
where $ c_{eff}=c+2\epsilon(3-2 \epsilon)(\gamma/\beta)$ and
$m_{\phi eff}$ is the mass of the amplitude and $m_{\theta eff}$ is
that of the phase. For $t>t_1$, $\psi$ decays, so the minimum and the
mass are given by 
\begin{equation}
\theta_0 = 0,m_{\ast \theta}^2 =0, \bar{\phi}_{0} = \left
  ( \frac{\sqrt{c}H  M}
{\sqrt{3}}\right)^{\frac{1}{2}},m_{\ast \phi}^2 = 4c.
\label{eqn:b2}
\end{equation} 
For $t<t_1$, the curvature around the minimum is large so fluctuations 
are massive ($m_{eff}>m$). 
Then, the amplitudes of the modes which exit 
the horizon for $t < t_1$ decrease with time more rapidly after crossing the 
horizon than the modes which exit the horizon for $t>t_1$. 
So, the decay of $\psi$ for $t>t_1$ gives the break to the
spectrum at $k_1$ which is the wavenumber of 
the mode which exits the horizon at $t=t_1$. The modes of $k<k_1$ have 
the steeper spectrum than the modes of $k>k_1$. 
Taking into consideration that the Hubble parameter varies with 
time \cite{Stewart}, we obtain the mode functions of the modes $k>k_1$ at 
the end of the inflation  
\begin{equation}
\left|\bar{\phi}_0 \delta \theta_k |_{H_e} \right| \sim  \frac{H_1}
{\sqrt{2 k^3}}
\left( \frac{k}{k_1} \right)^{\frac{3}{2}-\nu_{\theta}} \left(\frac{R_e}{R_1} 
\right)^{-q_{\theta}},    
\label{b3}
\end{equation} 
where
\begin{displaymath} 
q_{\theta} = \frac{(3-\epsilon)-\sqrt{(3-\epsilon)^2-4 m_{\ast \theta}^2}}{2},
\nu_{\theta} = \frac{\sqrt{(3-\epsilon)^2-4 m_{\ast \theta}^2}}{2(1-\epsilon)},
\nonumber  
\end{displaymath}
and $H_1$ and $R_1$ are the Hubble parameter and the scale factor when the
 mode with wavenumber $k_1$ exits the horizon.  The mode functions of the 
modes $k<k_1$ are obtained by replacing $\nu_{\theta}$ to $\nu_{\theta eff}$ 
which is given by replacing $m_{\ast \theta} $ to $m_{\ast \theta eff}$
Notice that the break of the spectrum is $k_1$ rather than the Hubble 
parameter during inflation as required. The spectrum index 
$n=-2\nu_{\theta eff}$ 
for $k<k_1$ is larger than $n=-2\nu_{\theta}$ for $k>k_1$. 
Fluctuations of the amplitude are given by replacing $q_{\theta}$,
$\nu_{\theta}$ to $q_{\phi}$, $\nu_{\phi}$ respectively.

(2){\it Subsequent to inflation} ($H>m_{3/2}$), we assume the universe 
enters the matter era during inflaton oscillation, so the Hubble parameter 
varies with time. For $H >m_{3/2}$, we can still ignore the terms proportional 
to $m_{3/2}$. Since we assume $\bar{a}=0$, the phase and its fluctuations are 
constant ($\theta =\theta_0 , \: \delta \theta= \delta \theta |_{H_e}$). 
In the case $c$ is not so tiny, the field tracks the fixed point near 
the instantaneous minimum of the potential and decreases with time 
($\sim (H M)^{1/2} \propto t^{-1/2}$) \cite{Dine} and the fluctuations 
vary in the same way ($\propto t^{-1/2}$), thus during this times 
$\delta \bar{\phi}/\bar{\phi}_0=\mbox{const.}$. 
If $c$ is very tiny or vanishes \cite{Stewart2}, the higher terms becomes 
important, so the AD fields is near critically damped 
($V''(\phi) \sim H^2$). In this case there also exists the
attracting point ($\sim (H M)^{1/2}$), so the situation is almost 
the same with the former case \cite{Dine}.
    
(3){\it At $H \sim m_{3/2}$}, the low energy supersymmetry breaking terms 
proportional to $m_{3/2}$ become important and the $B-L$ non-conserving 
A-term arises. The AD field falls toward the origin. The evolution of the 
AD field at this epoch can be considered as the motion of a point-like 
particle in the two dimensional plane $(\phi_R,\phi_I)$, where $\phi_R$ 
and $\phi_I$ are the real and imaginary part of the AD field. The baryon 
charge $n_B =2 | \bar{\phi} |^2 \dot{\theta}= 2(\dot{\phi_I} 
\phi_R-\dot{\phi_R} \phi_I)$ corresponds to the angular momentum of the 
particle, and its non-conservation is caused by $U(1)$-asymmetric A-term 
\cite{Kirilova}. In the case $\bar{A}=c= \eta \ll 1 $, we can 
obtain the approximate solution. The ansatz of the solution is
\begin{eqnarray}
\phi_R &=& \frac{t_i}{t} \alpha_R \sin(t-t_i +\beta_R 
(\frac{1}{t}-\frac{1}{t_i})+\delta_R) ,\nonumber\\
\phi_I &=& \frac{t_i}{t} \alpha_I \; \sin(t-t_i +\beta_I 
(\frac{1}{t}-\frac{1}{t_i})+\delta_I),
\end{eqnarray}
where we rescale $\phi \to (m_{3/2}M)^{1/2} \phi$ and $t \to m_{3/2}^{-1}t$.
The initial condition is set at $t=t_i$. Until 
$t=t_i$, we assume the AD fields tracks the fixed point, so at that time 
$\bar{\phi}=\bar{\phi}_i,\: \dot{\bar{\phi}}=-(1/2 t_i)
\bar{\phi},\: \theta=\theta_0, \: \dot{\theta}=0$. $\bar{\phi}_i$ is
determined by the time when the baryon number is excited ($t=t_i$). 
At this time  $H \sim m_{3/2}$, so $\bar{\phi}_i$ is $\sim O(1)$.
We obtain the solution to the first order of $\eta$
\begin{eqnarray}
\alpha_R &=& \frac{\sqrt{5}}{2} \phi_{Ri} ,\:\: \beta_R=\frac{\eta}{2} 
\left(\frac{4}{9}-\frac{15}{16} (\phi_{Ri}^2-3 \phi_{Ii}^2) 
\right),
\nonumber\\
\alpha_I &=& \frac{\sqrt{5}}{2} \phi_{Ii} \: ,\:\: \beta_I= 
\frac{\eta}{2} \left(\frac{4}{9}-\frac{15}{16}(\phi_{Ii}^2-3 
\phi_{Ri}^2)\right),
\end{eqnarray}
and $\delta_R \!=\delta_I \!=\arctan(2)$, where $\phi_{Ri}=\bar{\phi}_i
\cos(\theta_A/4), \phi_{Ii}=\bar{\phi}_i\sin (\theta_A/4)$.
Here we neglect the terms proportional to $t^p \;(p<-2)$ and the high 
frequency terms. The baryon number at late times tends to
\begin{equation}
n_B \sim \frac{75}{64} \left(\frac{t_i}{t} \right)^2 (m_{3/2}^2 M) \bar{A} 
\: \bar{\phi_i}^4 \sin\theta_A.
\end{equation}
We have done numerical calculations and find that this approximate solution 
agrees with numerical results up to a factor $2$. We consider fluctuations 
of the field as fluctuations in the initial condition of this motion. 
For $H >m_{3/2}$, $\delta \theta$ and $\delta \bar{\phi}/\bar{\phi_0}$ 
are approximately constant, 
so fluctuations in the baryon number become
\begin{equation}
\frac{\delta n_B}{n_B} \sim 4 \left( \cot \theta_A \left.\delta 
\theta \right |_{H=H_e} + \left. \frac{\delta \bar{\phi}}
{\bar{\phi}_0} \right |_{H=H_e} \right).
\label{c1} 
\end{equation}
In the case $\bar{a}=0,c\ne0$, fluctuations of the amplitude are suppressed by
the factor $(R_e/R_1)^{-q_{\phi}}$ compared to those of the phase so
negligible.  For $H< m_{3/2}$ A-term becomes negligible, so the baryon 
number and its fluctuations are constant in the comoving volume.

(4){\it After reheating}, the inflaton $\chi$ converts to the radiation and 
the entropy is produced. The fractional energy in the AD condensate at 
this time is $ \rho_{\phi}/\rho_{I} \sim (m_{3/2}M)/M_p^2 \ll 1 $, so
the entropy is dominated by inflaton decay. To ensure the condensate
is not damped by thermal scattering before the baryon asymmetry is
established, the reheating temperature should be $T_R<10^6$ Gev
\cite{Dine}. For the reheating temperature above the weak scale, $B-L$ is 
converted to the baryon number. 

We will show that both the appropriate baryon to entropy ratio and its 
fluctuations with the appropriate amplitude and spectrum shape are obtained 
by choosing specific parameters. The baryon to entropy ratio is \cite{Dine}  
\begin{equation}
\!\!\!\!\!\!\! \frac{n_B}{s} \sim  \frac{75}{64} \bar{A} \: \bar{\phi}_i^4 
\sin \theta_A \:  10^{-10} \left(\frac{T_R}{10^6 \mbox{Gev}} \right)
\left(\frac{10^{-3}M}{M_p} \right).
\end{equation}
When we choose $\bar{A}=c=0.1$, $\bar{\phi}_i=1,\theta_A=\pi/4$, 
$M=10^{3}M_p$, and $T_R=10^6 \mbox{Gev}$, this becomes $n_B/s=10^{-11}$, 
which is consistent with the constraint from primordial nucleosynthesis. 
The power spectrum of the entropy perturbation is given by 
\begin{equation}
\left \langle \left[ \frac{\delta(n_B /s)}{n_B /s} \right]_k^2 \right 
\rangle =Q \: P_B(k),
\end{equation}
where
\begin{eqnarray}
\quad Q &\sim& 10^{-3} \left(\frac{H_1}{M_p} \right) \left(\frac{H_1}{H_e} 
\right) \left(\frac{M_p}{10^{-3} M} \right)
\nonumber\\
k^3 P_B(k) &=& \left( \frac{k}{k_1} \right)^{3-2 \nu_{\theta}} \!\!\!\! 
(k \gg k_1),\: \left( \frac{k}{k_1} \right)^{3-2 \nu_{\theta eff}}\!\!\!\! 
(k \ll k_1). \nonumber 
\end{eqnarray}
We set the expansion rate during inflation at $\epsilon=0.28$. The rest 
parameters are determined so as to satisfy astrophysical constraints. 
First, in order to suppress the 
CMB anisotropy, the spectrum index should be $n>-1.5$ on large scales 
\cite{Efstathiou,Yokoyama}. So, we choose $\gamma/ \beta=0.57$, 
which corresponds to $m_{\ast \theta eff}^2=1.56$, then the spectrum 
index becomes $n=-2 \nu_{\theta eff}=-1.5$. The amplitude and the 
break are determined by $J_3$ normalization \cite{Yokoyama} 
\begin{eqnarray}
&&J_3(25 h^{-1} \mbox{Mpc})= \left( \frac{D_0}{D_{rec}}\right)^2 \int dk 
P_{rec}(k) \frac{\sin kr-kr \cos kr}{k} \nonumber\\
&&= 780 h^{-3} \mbox{Mpc}^3 \left(\frac{Q}{10^{-3}} \right) 
\left(\frac{D_0/D_{rec}}{90} \right)^2 \left(\frac{x_1}{\mbox{1Mpc}}\right)^{1.5} 
,
\end{eqnarray}
where $x_1=2 \pi/k_1$ is the comoving scale of the break, $D_0/D_{rec}$ 
is the growth factor from recombination to present and $P_{rec}$ is the 
power spectrum at recombination; $P_{rec} \!\!=\!\! P_B \;
(k \! \gg \! k_{eq}),\: P_{rec} \!\!=\!\! P_B (k/k_{eq})^4 \;(k \! \ll \! 
k_{eq})$. Here $k_{eq}$ is the comoving wavenumber corresponding to the 
horizon scale at the time of equality between baryon and radiation density. 
For example, we fix the break at $x_1=1 \mbox{Mpc}$ \cite{Peebles2}, 
then the amplitude $Q$ should be $10^{-3}$. To obtain the appropriate 
baryon to entropy ratio, $H_e$ should be larger than $m_{3/2}$. So we 
take $H_e=10^{5}\mbox{Gev}$ and $H_1 = 10^{12} \mbox{Gev}$, then 
$Q \sim 10^{-3}$. The expansion factor during inflation is given by 
\begin{equation}
\frac{R_1}{R_p}=\left(\frac{H_1}{H_e}\right)^{1/\epsilon} =10^{25}. 
\end{equation}
This is appropriate to adjust the scale of $k_1$ to $1 \mbox{Mpc}$.  
The value of the Hubble parameter during inflation is relatively low, so 
the primeval adiabatic perturbation is naturally subdominant. In fact, 
the curvature perturbation ${\cal R}$ of the inflaton $\chi$ can be 
estimated by the standard formula; 
\begin{equation}
{\cal R} \sim \frac{U(\chi)^{\frac{3}{2}}}{ M_p^3 U,_{\chi}} \sim \frac{1}
{\sqrt{16 \pi \epsilon}} \frac{H}{M_p} < 10^{-7},
\end{equation}
which is negligible compared to the baryon isocurvature perturbation. We 
make assumption that during inflation the coupling to the AD field does 
not affect the evolution of $\psi$ ($\beta \psi^2,\mu^2 \gg \gamma 
|\phi|^2$). Since the mass of $\psi$ is related to $H_1$ by $\mu^2=
\epsilon  H_1^2$ and $\psi^2 > \epsilon \beta^{-1}  H_1^2$ for $t<t_1$, 
this condition is satisfied if $\gamma < \epsilon H_1 /M$, which is 
consintent with the assumption $\gamma \sim \beta \sim m_{3/2}/M$ 

In summary, we have obtained the PBI, which gives both the appropriate 
baryon to entropy ratio $\sim 10^{-11}$ and the gaussian baryon number 
perturbation of the order $\sim 10^{-3}$ at the break $x_1=1 \mbox{Mpc}$ 
with the steep spectrum $n=-1.5$ on large scales $x>x_1$. The inflation 
occurs at relatively low scale such as $10^{12}\; \mbox{Gev}$, so the 
adiabatic perturbation generated by the inflaton is negligible naturally. 
The e-folding number during inflation is sufficient to solve the horizon 
problem. The reheating temperature is higher than the weak scale and low 
enough to avoid the gravitino problem.

The most essential assumption of our model is the coupling between AD field 
and the field $\psi$. The tiny coupling constant $\gamma  \sim m_{3/2}/M$ 
is needed \cite{super}. In our model, the coupling to $\psi$ produces 
the large initial values for baryogenesis in early stage of inflation, 
so even in the case when the terms coming from finite energy supersymmetry
breaking are small or vanishes \cite{Stewart2}, 
our model still works well. The exact value 
of coefficients of the potential relevant to the AD field and $\psi$ may 
be determined by high energy physics one has not known well and a set of 
parameters we take here is not unique. Our purpose in this letter is to 
show there exists at least one PBI in physical model of baryogenesis.  
 
The authors are grateful to anonymous referees for suggestions that
improved the presentation. The work of J.S. was supported by Monbusho 
Grant-in-Aid No.10740118.

\end{multicols}


\begin{references}
\bibitem{Exam} A simple example is the $H_{u}L$ direction which carries
 $B-L=-1$.
\bibitem{super}This potential is generated by the superpotential 
$W=(2 \sqrt{\beta^{-1}}\gamma \Phi^2- \sqrt{\beta}/2 \Psi^2)\Xi$ 
where $\Phi,\Psi,\Xi$ is the superfield of a scalar field
$\phi,\psi,\xi$. The AD field is no longer flat. However,
since $\gamma,\beta$ is the square of the yukawa coupling, 
these could be tiny [L.Randall, M.Soljacic and A.Guth, Nucl Phys.{\bf B472}, 
377 (1996)]. In particularly, if $\gamma \sim \beta 
\sim m_{3/2}/M \ll 1$, the quartic term of the AD field is always
negligible compared to the terms of eq.(\ref{eqn:b3}).
\end{references}
\end{document}